\documentclass[aps, prl, twocolumn, superscriptaddress,preprintnumbers]{revtex4-2}
\usepackage{bm, amsmath, amsfonts, amssymb, ascmac, mathtools, braket}
\usepackage{multirow}
\usepackage{graphicx}
\usepackage{float, color, xcolor}

\definecolor{rred}{rgb}{0.8, 0.0, 0.0}
\definecolor{bblue}{rgb}{0.0, 0.0, 0.8}

\usepackage[whole]{bxcjkjatype} % for Japanese
\usepackage{subfigure} % for subfigure
\usepackage{amscd} % for homological alg

\usepackage{bbm}
\usepackage{tabularx}

\usepackage{comment}

\usepackage[
pagebackref=false,
colorlinks=true,
linkcolor=bblue,
urlcolor=bblue,
filecolor=black,
citecolor=rred,
pdfstartview=FitV,
pdftitle={},
pdfauthor={},
pdfsubject={},
pdfkeywords={},
pdfpagemode=None,
bookmarksopen=true
]{hyperref}

\newcommand{\ii}{\text{i}}

\begin{document}

\title{Hopf Bifurcation of Nonlinear Non-Hermitian Skin Effect}

\author{Kohei Kawabata}
\email{kawabata@issp.u-tokyo.ac.jp}
\affiliation{Institute for Solid State Physics, University of Tokyo, Kashiwa, Chiba 277-8581, Japan}

\author{Daichi Nakamura}
\email{daichi.nakamura@issp.u-tokyo.ac.jp}
\affiliation{Institute for Solid State Physics, University of Tokyo, Kashiwa, Chiba 277-8581, Japan}

\date{\today}

\begin{abstract}
The non-Hermitian skin effect, nonreciprocity-induced anomalous localization of an extensive number of eigenstates, represents a hallmark of non-Hermitian topological systems with no analogs in Hermitian systems.
Despite its significance across various open classical and quantum systems, the influence of nonlinearity has remained largely unclear.
Here, we reveal the Hopf bifurcation of the nonlinear skin effect as a critical phenomenon unique to nonlinear non-Hermitian systems.
We demonstrate that nonlinearity destabilizes skin states and instead gives rise to the emergence of delocalized states associated with limit cycles in phase space. 
We also uncover the algebraically localized critical skin effect precisely at the Hopf bifurcation point.
We illustrate these behavior in a nonlinear extension of the Hatano-Nelson model in both continuum and lattice.
Our work shows a significant role of nonlinearity in the skin effect and uncovers rich phenomena arising from the interplay between non-Hermiticity and nonlinearity.
\end{abstract}

\maketitle

%%%%%%%%%%%%%%%%%%%%%%%%
Topological phases of matter constitute a cornerstone in condensed matter physics.
%Specifically, 
In band insulators and Bogoliubov-de Gennes superconductors, topology of wave functions has been extensively classified~\cite{HK-review, QZ-review, CTSR-review}.
The nontrivial bulk topology manifests itself in the emergence of anomalous gapless states at boundaries:
bulk-boundary correspondence.
Beyond quantum materials, topological band theory extends its applicability to a broad spectrum of classical synthetic platforms, including photonic~\cite{Lu-review, Ozawa-review} and mechanical~\cite{Huber-review, Ma-review} systems, as well as active matter~\cite{Shankar-review}.
Importantly, in such classical settings, nonlinearity is an inherent and pervasive feature. 
Accordingly, the impact of nonlinearity on topological insulators has attracted growing interest in recent years~\cite{Lumer-13, Chen-14, Leykam-16, ZangenehNejad-19, Mukherjee-20, Zhang-Wang-20, Tuloup-20, Maczewsky-20, Chaunsali-21, Zhou-22, Lo-21, Jurgensen-21, Mochizuki-21, Mostaan-22, Fu-22, Sone-24, Schindler-25, Isobe-24, Sone-25, Sone-Hatsugai-25, Smirnova-review, Szameit-review}.
For example, nonlinearity has been shown to induce solitons in topological insulators~\cite{Lumer-13, Chen-14, Leykam-16, Mukherjee-20, Zhang-Wang-20}.
It has also been demonstrated to quantize a topological pump through soliton formation and pitchfork or saddle-node bifurcations~\cite{Jurgensen-21, Mostaan-22, Fu-22}.

Meanwhile, topological characterization of non-Hermitian systems has also been studied considerably in both theory~\cite{Rudner-09, Sato-11, *Esaki-11, Hu-11, Schomerus-13, Longhi-15, Leykam-17, Xu-17, Shen-18, Takata-18, Gong-18, McDonald-18, KSUS-19, ZL-19, Wanjura-20, Denner-21, BBK-review, Okuma-Sato-review} and experiments~\cite{Zeuner-15, Zhen-15, *Zhou-18, Weimann-17, Xiao-17}.
Non-Hermiticity generally arises from the exchange of particles and energy with the external environment and gives rise to a variety of phenomena that have no counterparts in conservative systems~\cite{Konotop-review, Christodoulides-review}.
The bulk-boundary correspondence for non-Hermitian topology appears as the skin effect, anomalous localization of a macroscopic number of eigenstates driven by nonreciprocal dissipation~\cite{Lee-16, Xiong-18, MartinezAlvarez-18, YW-18-SSH, *YSW-18-Chern, Kunst-18, Lee-Thomale-19, Liu-19, Lee-Li-Gong-19, Herviou-19, Zirnstein-19, Borgnia-19, Yokomizo-19, Zhang-20, OKSS-20, Li-20, Okugawa-20, KSS-20, Zhang-22, Nakamura-24, Kawabata-23, Wang-24, Nakai-24, Zhang-24}.
Given that such an extraordinary sensitivity to boundary conditions is absent in Hermitian systems, the skin effect is a universal phenomenon intrinsic to non-Hermitian systems.
It has been observed across various open classical experiments of mechanical systems~\cite{Brandenbourger-19-skin-exp, *Ghatak-19-skin-exp}, electric circuits~\cite{Helbig-19-skin-exp, *Hofmann-19-skin-exp, Zhang-21}, photonic systems~\cite{Weidemann-20-skin-exp}, and active matter~\cite{Palacios-21}, in addition to open quantum ones of single photons~\cite{Xiao-19-skin-exp}, ultracold atoms~\cite{Liang-22, Zhao-25}, and digital processors~\cite{Shen-25}.

Notably, both nonlinearity and non-Hermiticity comprise important ingredients underlying the physics of open systems.
Prime recent examples include topological lasers, where the high-efficiency lasing is accomplished through the interplay of non-Hermiticity, topology, and nonlinearity~\cite{Poli-15, St-Jean-17, Parto-17, Bahari-17, Zhao-18, Harari-18, *Bandres-18}.
Consequently, the fate of topological phases in the simultaneous presence of nonlinearity and non-Hermiticity has recently been investigated~\cite{Xia-21, Pernet-22, Yuce-21, *Yuce-25, Ezawa-22, Zhu-22, Martello-23, Dai-24, ManyManda24HN, *ManyManda24soliton, Veenstra-24, Yoshida-25, Wang-25, Longhi-25}.
Nonlinear non-Hermitian eigenvalue problems are also relevant to the Green's function formalism at equilibrium~\cite{Kozii-17} and boost deformations~\cite{Nakai-22, *Guo-23}.
The combination of the skin effect and nonlinearity was shown to induce the single-mode lasing~\cite{Zhu-22} and soliton formation~\cite{ManyManda24soliton, Wang-25}.
However, the role of nonlinearity on the non-Hermitian skin effect, along with associated unique nonlinear non-Hermitian phenomena, has remained largely unexplored.

In this Letter, we reveal the Hopf bifurcation of the nonlinear skin effect as a critical phenomenon unique to nonlinear non-Hermitian systems.
Introducing a nonlinear extension of the Hatano-Nelson model, a prototypical model exhibiting the skin effect in the linear regime~\cite{Hatano-Nelson-96, *Hatano-Nelson-97}, we demonstrate that nonlinearity destabilizes skin states and instead gives rise to the emergence of delocalized states.
We further show that this localization transition of the nonlinear skin effect corresponds to the Hopf bifurcation in the theory of dynamical systems~\cite{Strogatz-textbook, Hirsch-textbook}---a transition from a stable fixed point to a limit cycle in phase space.
Moreover, we uncover that the exponential localization of skin states transforms into the algebraic (i.e., power-law) localization precisely at the Hopf bifurcation point, serving as a hallmark of the criticality.
We also substantiate these behavior in a lattice counterpart of the nonlinear Hatano-Nelson model.
Our work elucidates a significant role of nonlinearity in the skin effect and provides distinctive dissipative phenomena arising from the interplay of non-Hermiticity and nonlinearity.

%%%%%%%%%%%%%%%%%%%%%%%%
{\it Nonlinear Hatano-Nelson model}.---We consider the nonlinear Schr\"odinger equation in one-dimensional continuum space,
\begin{equation}
    \hat{H} \left( \psi \right) \psi = E \psi,
        \label{eq: nonlinear Schrodinger}
\end{equation}
where $E$ and $\psi$ denote an eigenvalue and its eigenstate, respectively.
Specifically, we study a nonlinear extension of the Hatano-Nelson model~\cite{Hatano-Nelson-96, *Hatano-Nelson-97},
\begin{equation}
    \hat{H} \left( \psi \right) = \frac{\hat{k}^2}{2} + \ii \left( \gamma - \varepsilon \left| \psi \right|^2 \right) \hat{k},
        \label{eq: nonlinear Hatano-Nelson}
\end{equation}
with a momentum operator $\hat{k} \coloneqq - \ii \partial_x$.
Here, $\gamma \in \mathbb{R}$ ($\varepsilon \geq 0$) quantifies the degree of linear (nonlinear) non-Hermiticity.
The non-Hermitian term $\gamma > 0$ enhances right-propagating waves with positive momenta and suppresses left-propagating ones with negative momenta, 
manifesting as the asymmetry of the hopping dynamics in real space.
While this asymmetric hopping physically describes a tilted magnetic field in type-II superconductors for the original Hatano-Nelson model~\cite{Hatano-Nelson-96, *Hatano-Nelson-97}, it can also be implemented, for example, in active mechanical~\cite{Brandenbourger-19-skin-exp, *Ghatak-19-skin-exp}, electric~\cite{Helbig-19-skin-exp, *Hofmann-19-skin-exp, Zhang-21}, and photonic~\cite{Weidemann-20-skin-exp} systems.
Meanwhile, $\varepsilon$ characterizes the modulation of the asymmetric hopping that depends nonlinearly on the local wave function amplitude $\left| \psi \right|^2$.
Under the nonreciprocal operation $\hat{k} \mapsto -\hat{k}$, this model returns back to itself only if both $\gamma$ and $\varepsilon$ simultaneously change their signs.
We assume that the norm of wave functions is unconstrained, corresponding to open classical systems.

\begin{figure}[t]
\centering
\includegraphics[width=1.0\linewidth]{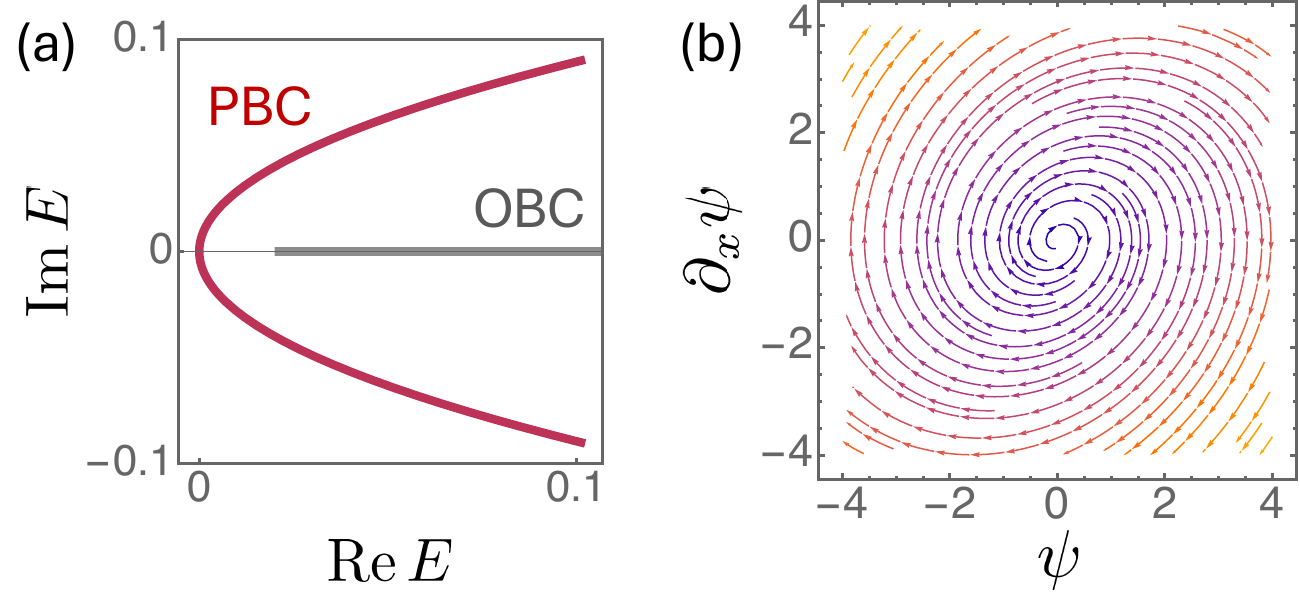} 
\caption{Linear non-Hermitian skin effect ($\varepsilon = 0$, $\gamma = 0.2$).
(a)~Complex spectra under the periodic boundary conditions (PBC; red) and open boundary conditions (OBC; gray).
(b)~Flow in phase space $\left( \psi, \partial_x \psi \right)$ ($E=0.5$).}	
    \label{fig: linear}
\end{figure}

%%%%%%%%%%%%%%%%%%%%%%%%
{\it Linear skin effect}.---In the linear regime $\varepsilon = 0$, non-Hermiticity $\gamma \neq 0$ causes the skin effect~\cite{Lee-16, YW-18-SSH, *YSW-18-Chern, Kunst-18, Yokomizo-19, Zhang-20, OKSS-20}.
Under the periodic boundary conditions with the system length $L > 0$, eigenenergies $E$ and the corresponding eigenstates $\psi$ are given as 
\begin{equation}
E \left( k \right) = \frac{k^2}{2} + \ii \gamma k, \quad \psi \left( x \right) \propto e^{\ii kx}
\end{equation}
with $k = 2\pi n/L$ ($n \in \mathbb{Z}$) [Fig.~\ref{fig: linear}\,(a)].
By contrast, in the presence of boundaries, the asymmetric hopping $\gamma \neq 0$ localizes eigenstates at the boundaries, thereby changing the eigenvalue problem drastically.
Indeed, under the open boundary conditions $\psi \left( 0 \right) = \psi \left( L \right) = 0$, Eq.~(\ref{eq: nonlinear Schrodinger}) reduces to the Hermitian eigenvalue problem $\partial_x^2 \tilde{\psi} + \left( 2E - \gamma^2 \right) \tilde{\psi} = 0$ for the transformed wave function $\tilde{\psi} \coloneqq e^{-\gamma x}\psi$.
Consequently, $E$ and $\psi$ are obtained as
\begin{equation}
    E \left( k \right) = \frac{k^2 + \gamma^2}{2}, \quad \psi \left( x \right) \propto e^{\gamma x} \sin kx
        \label{eq: linear dispersion}
\end{equation}
with $k = \pi n/L$ ($n \in \mathbb{N}_{>0}$).
Notably, the spectra exhibit substantial differences depending on the boundary conditions, which is a hallmark of the skin effect unique to non-Hermitian systems.
Accordingly, all the wave functions are localized toward the right (left) for $\gamma > 0$ ($\gamma < 0$).

\begin{figure}[t]
\centering
\includegraphics[width=1.0\linewidth]{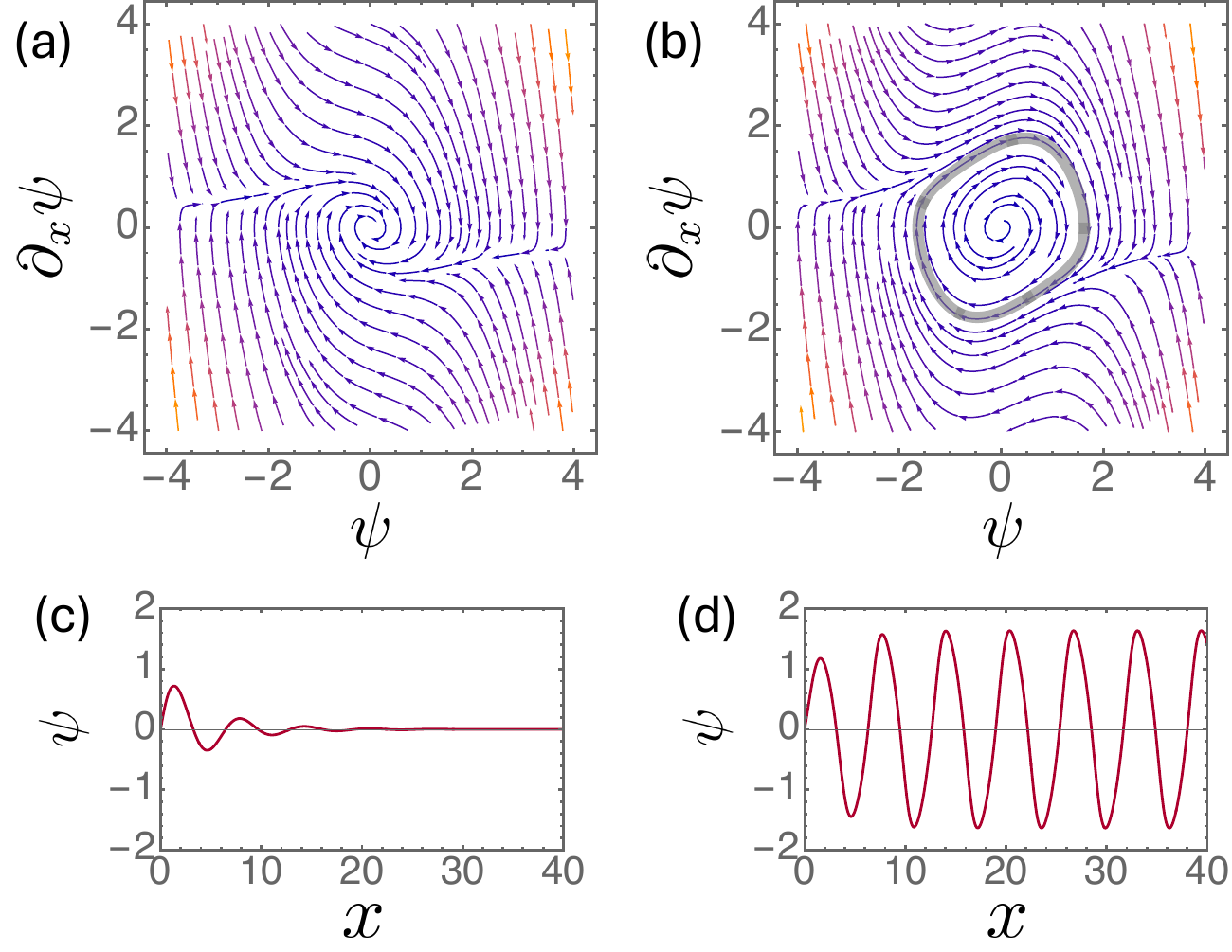} 
\caption{Hopf bifurcation in the nonlinear Hatano-Nelson model ($\varepsilon = 0.3$, $E = 0.5$) for (a, c)~$\gamma = - 0.2$ and (b, d)~$\gamma = 0.2$.
(a, b)~Flow in phase space $\left( \psi, \partial_x \psi \right)$.
The gray loop shows the limit cycle.
(c, d)~Wave function numerically obtained with the %initial 
boundary
conditions $\psi \left( x=0 \right) = 0$, $\psi' \left( x=0 \right) = 1$.}	
    \label{fig: continuum}
\end{figure}

%%%%%%%%%%%%%%%%%%%%%%%%
{\it Hopf bifurcation}.---We now proceed to study the influence of nonlinearity $\varepsilon > 0$.
As discussed earlier, the linear non-Hermitian term with $\gamma < 0$ induces an exponential decay of the wave function toward the right, making the nonlinear term $\varepsilon \left| \psi \right|^2$ negligible.
For $\gamma > 0$, by contrast, the linear non-Hermitian term amplifies the wave function whereas the nonlinear term gives a negative feedback mechanism.
Their competition can give rise to the formation of a delocalized state unique to nonlinear non-Hermitian systems.
We demonstrate that this is indeed the case and that nonlinearity destabilizes skin states and leads to delocalized states corresponding to a limit cycle in phase space.
Below, we focus on the case of $E > 0$ and $\psi \in \mathbb{R}$ to elucidate the influence of nonlinearity on linear skin states.
Notably, by replacing space with time, the nonlinear Schr\"odinger equation~(\ref{eq: nonlinear Schrodinger}) governing the nonlinear Hatano-Nelson model in Eq.~(\ref{eq: nonlinear Hatano-Nelson}) reduces to the equation of motion for the van der Pol oscillator~\cite{vanderPol-1926}---a prototype exhibiting a limit cycle.

Given the difficulty to analytically solve Eq.~(\ref{eq: nonlinear Schrodinger}) in the nonlinear regime, we introduce an approach of dynamical systems to elucidate the nonlinear non-Hermitian skin effect~\cite{Strogatz-textbook, Hirsch-textbook}.
We reformulate Eq.~(\ref{eq: nonlinear Schrodinger}) as
\begin{equation}
    \partial_x \begin{pmatrix}
        \psi \\ \partial_x \psi
    \end{pmatrix} = \begin{pmatrix}
        0 & 1 \\
        -2E & 2\,( \gamma - \varepsilon \left| \psi \right|^2 )
    \end{pmatrix} \begin{pmatrix}
        \psi \\ \partial_x \psi
    \end{pmatrix}
\end{equation}
and consider the associated flow in phase space $\left( \psi, \partial_x \psi \right)$.
We then investigate the spatial distribution of the wave function $\psi$ toward the right direction $x \geq 0$, corresponding to the semi-infinite boundary conditions, and study how the localization behavior depends on the eigenenergy $E$.
In the linear regime $\varepsilon = 0$, the system exhibits a unique fixed point at the origin $\left( \psi, \partial_x \psi \right) = \left( 0, 0 \right)$.
Around this fixed point, the eigenvalues of the coefficient matrix are 
\begin{equation}
    \lambda = \gamma \pm \sqrt{\gamma^2 - 2E}.
        \label{eq: eigs}
\end{equation}
For $\gamma < 0$ ($\gamma > 0$), both eigenvalues possess negative (positive) real parts, making the fixed point $\left( \psi, \partial_x \psi \right) = \left( 0, 0 \right)$ stable (unstable) [Fig.~\ref{fig: linear}\,(b)].
The wave function asymptotically behaves as $\psi \left( x \right) \sim e^{\lambda x}$ and converges to the origin (infinity) in phase space for $\gamma < 0$ ($\gamma > 0$), further implying the skin effect localized toward the left (right).
This constitutes the phase space analysis of the linear skin effect.

Even in the nonlinear regime, the origin $\left( \psi, \partial_x \psi \right) = \left( 0, 0 \right)$ continues to serve as the unique fixed point, which is characterized by the eigenvalues in Eq.~(\ref{eq: eigs}) and thus stable (unstable) for $\gamma < 0$ ($\gamma > 0$) [Fig.~\ref{fig: continuum}\,(a, b)].
A crucial distinction from the linear case is the emergence of a limit cycle for $\gamma > 0$:
a flow departing around the origin does not diverge to infinity but instead asymptotically approaches a closed periodic orbit.
Consequently, while the wave function decays toward the right for $\gamma < 0$ similar to the linear regime [Fig.~\ref{fig: continuum}\,(c)], it undergoes spatial delocalization for $\gamma > 0$ %, despite the presence of the asymmetric hopping 
[Fig.~\ref{fig: continuum}\,(d)].
This transition from a stable fixed point to a limit cycle represents the Hopf bifurcation in the framework of  dynamical systems.
We here unveil that the Hopf bifurcation manifests itself as the localization transition of the nonlinear skin effect.

%%%%%%%%%%%%%%%%%%%%%%%%
{\it Limit cycle}.---For arbitrary $\gamma > 0$, the existence and uniqueness of a limit cycle in phase space $\left( \psi, \partial_x \psi \right)$ are proved based on the Li\'enard theorem~\cite{Jordan-textbook, Grimshaw-textbook, Perko-textbook, supplement}.
Accordingly, delocalized eigenstates generally appear for $\gamma > 0$ even in the presence of the asymmetric hopping.
%Since their emergence is caused by the relative competition between $\gamma$ and $\varepsilon$, strong nonlinearity is not necessarily needed,
Hence, their emergence does not necessitate strong nonlinearity,
which enables a perturbative approach.
Indeed, in the weakly nonlinear regime [$0 < \varepsilon \ll 1$, $\gamma = \mathcal{O} \left( \varepsilon \right)$], the energy dispersion of these delocalized eigenstates is obtained as 
\begin{equation}
    E \left( k \right) \simeq \frac{k^2}{2} + \frac{\gamma^2}{4}
\end{equation}
on the basis of singular perturbation theory (see the End Matter and Supplemental Material~\cite{supplement} for details).
Notably, all the eigenstates with $E \geq \gamma^2/4$ are delocalized, in contrast to the skin effect.
The distinction from the linear case in Eq.~(\ref{eq: linear dispersion}) is also reflected in the additional constant term.
Conversely, in the strongly nonlinear regime ($\varepsilon \gg 1$), the energy dispersion is evaluated as
\begin{equation}
    E \left( k \right) \simeq \frac{3 - 2\log 2}{2\pi} \gamma k.
        \label{eq: strong nonlinearity}
\end{equation}
Despite the nonrelativistic nature of the original system, the strong nonlinearity induces the emergent relativistic energy dispersion.

%%%%%%%%%%%%%%%%%%%%%%%%
{\it Critical nonlinear skin effect}.---Exactly at the onset of the Hopf bifurcation, algebraically localized skin states emerge.
For $\gamma = 0$, the characteristic length scale vanishes, implying a signature of criticality.
When the wave function is expanded as $\psi \left( x \right) \sim \phi \left( x \right) \sin\,( \sqrt{2E} x )$ for $x \to \infty$, the nonlinear Schr\"odinger equation~(\ref{eq: nonlinear Schrodinger}) asymptotically reduces to 
\begin{equation}
    \partial_x \phi \sim - \varepsilon \phi^3 \sin^2\,( \sqrt{2E} x ) \sim - \frac{\varepsilon}{2} \phi^3,
\end{equation}
where the oscillatory term is averaged over space.
Consequently, the wave function exhibits the power-law decay 
\begin{equation}
\phi \left( x \right) \sim \frac{1}{\sqrt{\varepsilon x}}
    \label{eq: power law}
\end{equation}
for $x \to \infty$.
We confirm this critical behavior through numerically solving Eq.~(\ref{eq: nonlinear Schrodinger}) for $\gamma = 0$~\cite{supplement}.
While algebraically decaying skin states also appear in the linear regime~\cite{Li-20, Kawabata-23, Zhang-24}, the ones identified here arise from the interplay between nonlinearity and non-Hermiticity.

%%%%%%%%%%%%%%%%%%%%%%%%
{\it Point-gap topology}.---In the linear regime, the skin effect originates from point-gap topology~\cite{Zhang-20, OKSS-20}.
In the presence of a point gap, complex eigenenergies $E_n$'s are defined not to intersect a reference point $E_{\rm P}$ (i.e., $E_n \neq E_{\rm P}$)~\cite{Gong-18, KSUS-19}.
To characterize the point-gap topology, we introduce the winding number $W \left( E_{\rm P} \right)$ of the complex spectrum by
\begin{equation}
    W \left( E_{\rm P} \right) = - \oint \frac{dk}{2\pi\ii} \left( \frac{d}{dk} \log \det \left[ H \left( k \right) - E_{\rm P} \right] \right),
        \label{eq: winding number}
\end{equation}
where the integral is performed for entire momenta.
Since $W \left( E_{\rm P} \right)$ vanishes in Hermitian systems, nontrivial $W \left( E_{\rm P} \right) \neq 0$ provides a topological invariant unique to non-Hermitian systems.
For the Hatano-Nelson model in Eq.~(\ref{eq: nonlinear Hatano-Nelson}) without nonlinearity (i.e., $\varepsilon = 0$), we have $W \left( E_{\rm P} \right) = \mathrm{sgn}\,\gamma$ for the reference energy $E_{\rm P} > 0$.
This nontrivial complex-spectral winding number under the periodic boundary conditions directly leads to the skin effect under the open boundary conditions, thereby establishing the bulk-boundary correspondence in non-Hermitian systems.

Even in the nonlinear regime $\varepsilon \neq 0$, plane waves $\psi \left( x \right) = a e^{\ii kx}$ persist as eigenstates with the eigenenergies $E \left( k \right) = k^2/2 + \ii\,( \gamma - \varepsilon \left| a \right|^2 )\,k$ under the periodic boundary conditions.
As a result, a point gap persists except for $\left| a \right|^2 \neq \gamma/\varepsilon$, in which case the winding number is determined as $W \left( E_{\rm P} \right) = \mathrm{sgn}\,( \gamma - \varepsilon \left| a \right|^2 )$ for $E_{\rm P} > 0$.
This holds even for $\gamma > 0$, where nonlinearity destabilizes skin states and leads to the emergence of delocalized states characterized by limit cycles.
Thus, the correspondence between the point-gap topology and skin effect seems to break down in the nonlinear regime.
Nevertheless, the emergence of delocalized eigenstates associated with limit cycles may still be related to point-gap topology in nonlinear non-Hermitian systems.
It should also be noted that nonlinearity can give rise to additional eigenstates beyond plane waves in contrast to the linear regime.
Such additional eigenstates may restore the bulk-boundary correspondence for point-gap topology, which is left for future studies.

\begin{figure}[t]
\centering
\includegraphics[width=1.0\linewidth]{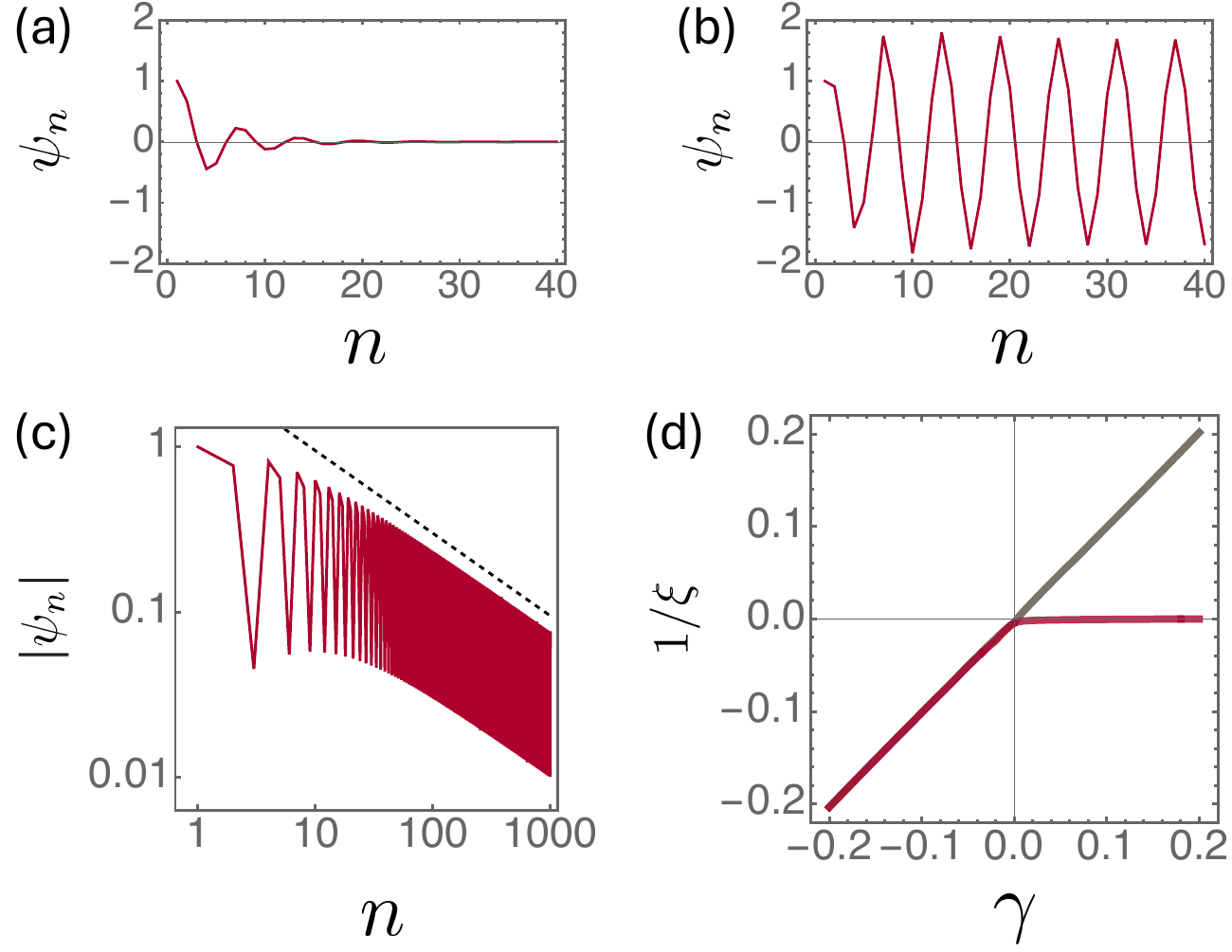} 
\caption{Nonlinear Hatano-Nelson model on a lattice ($E = 0.5$). 
%Initial 
Boundary
conditions: $\psi_0 = 0$, $\psi_1 = 1$.
(a-c)~Wave function for $\varepsilon = 0.3$ and (a)~$\gamma = -0.2$, (b)~$\gamma = 0.2$, and (c)~$\gamma = 0$ (dotted line: $\left| \psi_n \right| = 3/\sqrt{n}$).
(d)~Inverse of the localization length, $1/\xi$ ($\left| \psi_n \right| \sim e^{n/\xi}$), numerically obtained over $500 \leq n \leq 1000$ [$\varepsilon = 0$ (gray), $\varepsilon = 0.3$ (red)].}	
    \label{fig: lattice}
\end{figure}

%%%%%%%%%%%%%%%%%%%%%%%%
{\it Lattice model}.---We further elucidate the Hopf bifurcation of the nonlinear non-Hermitian skin effect in a lattice counterpart of Eq.~(\ref{eq: nonlinear Hatano-Nelson}).
A discrete counterpart of the nonlinear Schr\"odinger equation on a one-dimensional lattice is given by $\hat{H} \left( \ket{\psi} \right) \ket{\psi} = E \ket{\psi}$ with
\begin{align}
    &\hat{H} \left( \ket{\psi} \right) = \sum_{n} \left( - \frac{1 - \gamma + \varepsilon \left| \psi_n \right|^2}{2} \ket{n} \bra{n+1}
    \right. \nonumber \\ 
    &\qquad \left. - \frac{1 + \gamma - \varepsilon \left| \psi_{n+1} \right|^2}{2} \ket{n+1} \bra{n}
    + \ket{n} \bra{n} \right),
        \label{eq: NL HN lattice}
\end{align}
where $\ket{\psi} = \left( \psi_1, \cdots, \psi_L \right)^T$ represents a single-particle eigenstate, and $\ket{n}$ denotes a single-particle state localized at site $n$.
Similar to the continuum counterpart, $\gamma \in \mathbb{R}$ characterizes the degree of asymmetric hopping, and $\varepsilon \geq 0$ quantifies the strength of its nonlinear modulation that can induce a negative feedback mechanism for $\gamma > 0$.
Such a tight-binding model can be implemented, for example, in a laser-written array of waveguides, where individual elements are coupled with each other evanescently~\cite{Szameit-review}.

To investigate the localization behavior under the semi-infinite boundary conditions, we here employ the transfer matrix method~\cite{Kramer-10}.
We reformulate the nonlinear Schr\"odinger equation with the nonlinear transfer matrix $T \left( \psi_n \right)$:
\begin{align}
    &\begin{pmatrix}
        \psi_{n+1} \\ \psi_{n}
    \end{pmatrix} = T \left( \psi_n \right) \begin{pmatrix}
        \psi_{n} \\ \psi_{n-1}
    \end{pmatrix}, \\
    &T \left( \psi_n \right) \coloneqq \begin{pmatrix}
        - \cfrac{2 \left( E-1 \right)}{1-\gamma+\varepsilon \left| \psi_n \right|^2} & - \cfrac{1+\gamma-\varepsilon \left| \psi_n \right|^2}{1-\gamma+\varepsilon \left| \psi_n \right|^2} \\
        1 & 0
    \end{pmatrix}.
\end{align}
The spatial distributions of wave functions $\ket{\psi}$ for given energy $E$ are thus determined through the cumulative product of $T \left( \psi_n \right)$'s:
\begin{equation}
    \begin{pmatrix}
        \psi_{L+1} \\ \psi_{L}
    \end{pmatrix} = \left[ \prod_{n=1}^{L} T \left( \psi_n \right) \right] \begin{pmatrix}
        \psi_{1} \\ \psi_{0}
    \end{pmatrix} \eqqcolon T_L \begin{pmatrix}
        \psi_{1} \\ \psi_{0}
    \end{pmatrix}.
\end{equation}
Specifically, the localization length is characterized by $1/\xi = \lim_{L\to \infty} \alpha_L
/L$, where $\alpha_{L}$ denotes the smallest positive eigenvalue of $\log \sqrt{ T_L^{\dag} T_L }$.
A similar nonlinear transfer matrix approach has recently been used in the Hermitian setting~\cite{Sone-25, Sone-Hatsugai-25}.

We numerically calculate wave functions and their localization properties under the boundary conditions $\psi_0 = 0$ and $\psi_1 = 1$ (Fig.~\ref{fig: lattice}).
Consistent with the continuum counterpart, the wave functions are localized toward the left for $\gamma < 0$ but become delocalized for $\gamma > 0$.
Precisely at the Hopf bifurcation point $\gamma = 0$, the wave function exhibits the algebraic decay $\left| \psi_n \right| \propto 1/\sqrt{n}$, in agreement with Eq.~(\ref{eq: power law}).
Figure~\ref{fig: lattice}\,(d) shows the localization length $\xi$ as a function of linear non-Hermiticity $\gamma$.
In the linear regime (i.e., $\varepsilon = 0$), we obtain $\xi = 1/\gamma$, indicating the conventional skin effect.
In the nonlinear regime, on the other hand, while the skin effect persists for $\gamma < 0$, the localization length diverges, $\xi = \infty$, for $\gamma > 0$, implying the emergence of delocalized states.
These results demonstrate the Hopf bifurcation and accompanying delocalization of skin states even in the lattice model.
The Hopf bifurcation also manifests in the nonlinear dynamics, providing a clear experimental signature:
whereas a wave packet propagates unidirectionally in the linear regime, nonlinearity causes the bidirectional propagation even in the presence of the asymmetric hopping, reflecting the nonlinearity-induced destabilization of the skin effect and the concomitant emergence of delocalized states (see the End Matter for details).

%%%%%%%%%%%%%%%%%%%%%%%%
{\it Discussion}.---In this Letter, we have elucidated the Hopf bifurcation of the nonlinear skin effect as a phase transition induced by the interplay of non-Hermiticity and nonlinearity.
Since this phase transition is governed by the relative competition between linear and nonlinear non-Hermitian terms, strong nonlinearity is not strictly required, thereby potentially facilitating its experimental implementation.
While we have focused on a nonlinear extension of the Hatano-Nelson model in this Letter, the Hopf bifurcation is a ubiquitous phenomenon in nonlinear non-Hermitian systems, driving phase transitions in the nonlinear skin effect.
Indeed, we further demonstrate it in a nonlinear non-Hermitian extension of the Su-Schrieffer-Heeger model~\cite{SSH-79, supplement}.
Further exploration of the implications of other types of bifurcation for the nonlinear skin effect is warranted.

Moreover, the effect of disorder merits further investigation.
We find the coexistence of delocalized and localized eigenstates in the lattice version of the nonlinear Hatano-Nelson model in Eq.~(\ref{eq: NL HN lattice})~\cite{supplement}.
This shows the emergence of nonlinearity-induced mobility edges even in the absence of disorder, reminiscent of those found in the disordered Hatano-Nelson model~\cite{Hatano-Nelson-96, *Hatano-Nelson-97}.
The combination of disorder and nonlinearity can form distinctive mobility edges.
Developing a unified understanding of the localization transitions in nonlinear non-Hermitian systems can also bear practical significance for disorder-resilient wave transport phenomena.
Additionally, the skin effect can be enriched in different symmetry classes or higher dimensions in the linear regime~\cite{OKSS-20, Okugawa-20, KSS-20, Nakamura-24}.
Understanding the influence of nonlinearity therein is worth studying.
Finally, the nonlinearity-induced instability of the skin effect implies that nonlinearity can fundamentally change the topological classification and bulk-boundary correspondence.

\medskip
%%%%% Acknowledgement %%%%%
\begingroup
\renewcommand{\addcontentsline}[3]{}% Remove functionality of \addcontentsline
\begin{acknowledgments}
We thank Zhao-Fan Cai, Naomichi Hatano, Tomi Ohtsuki, and Zhenyu Xiao for helpful discussion.
K.K. thanks Kazuki Yamamoto for letting him know about StreamPlot of Mathematica.
K.K. is supported by MEXT KAKENHI Grant-in-Aid for Transformative Research Areas A ``Extreme Universe" No.~JP24H00945.
D.N. is supported by JSPS KAKENHI Grant No.~JP24K22857.
\end{acknowledgments}
\endgroup

\let\oldaddcontentsline\addcontentsline% Store \addcontentsline
\renewcommand{\addcontentsline}[3]{}% Make \addcontentsline a no-op
\bibliography{NH_top.bib}
\let\addcontentsline\oldaddcontentsline% Restore \addcontentsline

\section{End Matter}

\appendix
%%%%%%%%%
\setcounter{secnumdepth}{2}

%%%%%%%%%%%%
\section{Singular perturbation theory for weak nonlinearity}
    \label{endsec: weak}

Employing singular perturbation theory for weak nonlinearity $0 \leq \varepsilon \ll 1$, 
we derive a delocalized eigenstate associated with a limit cycle in phase space.
Here, we focus on the vicinity of the Hopf bifurcation and assume that the linear non-Hermitian parameter $\gamma$ is of the same order as the nonlinear non-Hermitian parameter $\varepsilon$, i.e., 
\begin{equation}
    \tilde{\gamma} \coloneqq \frac{\gamma}{\varepsilon} = \mathcal{O} \left( 1 \right).
\end{equation}
Under this assumption, the nonlinear Schr\"odinger equation reads
\begin{equation}
    \partial_x^2 \psi + 2 \varepsilon \left( \psi^2 - \tilde{\gamma} \right) \partial_x \psi + 2E \psi = 0.
        \label{endeq: Schrodinger weak nonlinear}
\end{equation}
To systematically capture the multiscale nature of the problem, we expand the wave function as
\begin{equation}
    \psi = \psi_0 + \varepsilon \psi_1 + \mathcal{O} \left( \varepsilon^2 \right)
\end{equation}
and introduce the two different length scales
\begin{equation}
    x_0 \coloneqq x, \quad x_1 \coloneqq \varepsilon x. 
\end{equation}
Given that
\begin{equation}
    \partial_x %= \left( \partial_x x_0 \right) \partial_{x_0} + \left( \partial_x x_1 \right) \partial_{x_1} %\nonumber \\
    = \partial_{x_0} + \varepsilon \partial_{x_1} %\nonumber \\
    \eqqcolon \partial_0 + \varepsilon \partial_1,
\end{equation}
we have 
\begin{align}
    \partial_x \psi &%= \left( \partial_0 + \varepsilon \partial_1 \right) \left( \psi_0 + \varepsilon \psi_1 \right) 
    \simeq \partial_0 \psi_0 + \varepsilon \left( \partial_1 \psi_0 + \partial_0 \psi_1 \right), %+ \varepsilon^2 \left( \partial_2 \psi_0 + \partial_1 \psi_1 + \partial_0 \psi_2 \right) 
    \\
    \partial_x^2 \psi &%= \left( \partial_0^2 + 2 \varepsilon \partial_0 \partial_1 
    %\right) \left( \psi_0 + \varepsilon \psi_1 \right) 
    \simeq \partial_0^2 \psi_0 + \varepsilon \left( 2\partial_0\partial_1 \psi_0 + \partial_0^2 \psi_1 \right). %+ \varepsilon^2 \left( \left( \partial_1^2 + 2\partial_0 \partial_2 \right) \psi_0 + 2\partial_0 \partial_1 \psi_1 +  \partial_0^2 \psi_2 \right)
\end{align}
Since Eq.~(\ref{endeq: Schrodinger weak nonlinear}) must hold at $\mathcal{O} \left( 1 \right)$ and $\mathcal{O} \left( \varepsilon \right)$ independently, we obtain
\begin{align}
    &\partial_0^2 \psi_0 + 2E \psi_0 = 0, 
        \label{aeq: NL1} \\
    &\partial_0^2 \psi_1 + 2E \psi_1 = -2 \partial_0 \partial_1 \psi_0 - 2 \left( \psi_0^2 - \tilde{\gamma} \right) \partial_0 \psi_0.   
        \label{aeq: NL2} %\\
    %&\partial_0^2 \psi_2 + 2E \psi_2 = - \left( \partial_1^2 + 2\partial_0 \partial_2 \right) \psi_0 - 2 \partial_0 \partial_1 \psi_1 - 2 \left( 2 \psi_0 \psi_1 \left( \partial_0 \psi_0 \right) + \left( \psi_0^2 - \tilde{\gamma}^2 \right) \left( \partial_1 \psi_0 + \partial_0 \psi_1 \right) \right)
        %\label{eq: NL3}
\end{align}

The general solution to Eq.~(\ref{aeq: NL1}) is given by
\begin{equation}
    \psi_0 = r_0 \sin \theta_0, \quad \theta_0 \coloneqq \sqrt{2E} x_0 + \phi_0,
\end{equation}
where $r_0 \geq 0$ and $\phi_0 \in \left[ 0, 2\pi \right)$ are functions of $x_1$ determined by the boundary conditions.
Substituting this solution into Eq.~(\ref{aeq: NL2}), we obtain
\begin{align}
    &\partial_0^2 \psi_1 + 2E \psi_1 = - 2 \sqrt{2E} \left( \left( \partial_1 r_0 \right) \cos \theta_0 - r_0 \left( \partial_1 \phi_0 \right) \sin \theta_0 \right) \nonumber \\
    &\qquad\qquad\qquad - 2 \sqrt{2E} r_0 \left( r^2_0 \sin^2 \theta_0 - \tilde{\gamma} \right) \cos \theta_0.
\end{align}
From the formula
\begin{equation}
    \sin^2 \theta_0 \cos \theta_0 = \frac{1}{4} \left( \cos \theta_0 - \cos 3\theta_0 \right),
\end{equation}
the conditions for the elimination of resonance terms yield
\begin{align}
    &\partial_1 r_0 = -r_0 \left( \frac{r_0^2}{4} - \tilde{\gamma} \right), 
        \label{aeq: NL2-1} \\
    &r_0 \left( \partial_1 \phi_0 \right) = 0.
        \label{aeq: NL2-2}
\end{align}
From Eq.~(\ref{aeq: NL2-1}), the existence of a limit cycle is dictated by the sign of $\gamma$:
\begin{itemize}
    \item For $\gamma < 0$, the fixed point $r_0 = 0$ is unique and stable.
    As a result, we have $r_0 \to 0$ for $x \to \infty$, implying the absence of limit cycles.

    \item For $\gamma > 0$, the fixed point $r_0 = 0$ becomes unstable, giving way to the stable fixed point $r_0 = 2\sqrt{\tilde{\gamma}}$.
    As a result, we have $r_0 \to 2\sqrt{\tilde{\gamma}}$ for $x \to \infty$, implying the emergence of a limit cycle.
\end{itemize}
Moreover, Eq.~(\ref{aeq: NL2-2}) indicates that the limit cycle satisfies $\partial_1 \phi_0 = 0$, further implying that $\phi_0$ remains a constant with respect to $x_1$.
Therefore, the limit cycle for $\gamma > 0$, corresponding to a delocalized eigenstate in real space, is described by
\begin{equation}
    \psi_0 \simeq 2\sqrt{\frac{\gamma}{\varepsilon}} \sin \left( k x + \mathcal{O} \left( \varepsilon^2 \right) \right), \quad k \coloneqq \sqrt{2E}
\end{equation}
for $x\to \infty$.
Equation~(\ref{aeq: NL2}) further reduces to
\begin{equation}
    \partial_0^2 \psi_1 + 2E\psi_1 %= \frac{\sqrt{2E}}{2} r_0^3 \cos 3\theta_0 
    = 4 \tilde{\gamma}^{3/2}
    \sqrt{2E} \cos 3\theta_0, 
\end{equation}
which admits the solution
\begin{equation}
    \psi_1 = r_1 \sin \theta_1 - \frac{\tilde{\gamma}^{3/2}}{\sqrt{8E}} \cos 3\theta_0, \quad \theta_1 \coloneqq\sqrt{2E} x_0 + \phi_1
\end{equation}
with $r_1 \geq 0$ and $\phi_1 \in \left[ 0, 2\pi \right)$ dependent on $x_1$.
In the Supplemental Material~\cite{supplement}, we further derive the $\mathcal{O} \left( \varepsilon^2 \right)$ correction to the energy dispersion using the Poincar\'e-Lindstedt method.

In passing, the leading-order wave function for $\gamma < 0$ is obtained from Eq.~(\ref{aeq: NL2-1}) as
\begin{equation}
    \psi_0 \simeq 2\sqrt{- \frac{\gamma}{\varepsilon}} \frac{e^{\gamma \left( x-4c \right)}}{\sqrt{1-e^{2\gamma \left( x- 4c \right)}}} \sin kx, \quad k \coloneqq \sqrt{2E},
\end{equation}
where the constant $c$ is determined by the boundary conditions.
This shows the nontrivial correction of skin states induced by nonlinearity.
Specifically, $\psi_0$ contains not only original skin states $e^{\gamma x}$ but also additional decaying contributions $e^{\left( 2n+1 \right) \gamma x}$ ($n \in \mathbb{N}_{>0}$), which constitutes a signature of the enhanced localization due to nonlinearity.

%%%%%%%%%%%%
\section{Limit cycle for strong nonlinearity}
    \label{endsec: strong}

We analyze a delocalized eigenstate associated with a limit cycle in phase space under the regime of strong nonlinearity $\varepsilon \gg 1$.
%Given that
%\begin{equation}
%    \partial_x^2 \psi + 2 \left( \varepsilon \psi^2 - \gamma \right) \partial_x \psi = \partial_x \left( \partial_x \psi + 2\varepsilon \left( \frac{1}{3} \psi^3 - \frac{\gamma}{\varepsilon} \psi\right) \right),
%\end{equation}
The nonlinear Schr\"odinger equation for the nonlinear Hatano-Nelson model reads
\begin{align}
    \partial_x \psi &= \varepsilon \left( \chi - 2F \left( \psi \right) \right), 
        \label{aeq: strong1} \\
    \partial_x \chi &= - \frac{2E}{\varepsilon} \psi,
        \label{aeq: strong2}
\end{align}
with
\begin{equation}
    F \left( \psi \right) \coloneqq \frac{1}{3} \psi^3 - \frac{\gamma}{\varepsilon} \psi, \quad \chi \coloneqq \frac{1}{\varepsilon }\partial_x \psi + 2F \left( \psi \right).
\end{equation}
Owing to $\varepsilon \gg 1$, Eq.~(\ref{aeq: strong1}) indicates that an arbitrary initial point in phase space $\left( \psi, \chi \right)$ rapidly converges to the nullcline $\chi = 2F \left( \psi \right)$.
Subsequently, it slowly drifts along this nullcline according to Eq.~(\ref{aeq: strong2}), ultimately leading to the convergence at the fixed point $\left( 0, 0 \right)$ for $\gamma \leq 0$ and the emergence of a limit cycle for $\gamma > 0$.
Since the cubic function $\chi = 2F \left( \psi \right)$ attains extrema $\pm \chi_0 \coloneqq \pm \left( 4/3 \right) \left( \gamma/\varepsilon \right)^{3/2}$ at $\psi = \pm \psi_0 = \pm \sqrt{\gamma/\varepsilon}$, the period of the limit cycle, corresponding to the wavelength of the delocalized eigenstate, is evaluated as
\begin{align}
    \lambda %= 2 \int_{2\psi_0}^{\psi_0} \frac{dx}{d\chi} \frac{d\chi}{d\psi} d\psi 
    \simeq 2 \int_{2\psi_0}^{\psi_0} \left( - \frac{\varepsilon}{2E\psi}\right) \left( 2F' \left( \psi \right) \right) d\psi 
    %= \frac{2\varepsilon}{E} \int_{\psi_0}^{2\psi_0} \left( \psi - \frac{\gamma}{\varepsilon} \frac{1}{\psi}\right) d\psi 
    = \left( 3 - 2 \log 2 \right) \frac{\gamma}{E}.
\end{align}
%\begin{align}
%    \lambda &= 2 \int_{2\psi_0}^{\psi_0} \frac{dx}{d\chi} \frac{d\chi}{d\psi} d\psi \nonumber \\
%    &\simeq 2 \int_{2\psi_0}^{\psi_0} \left( - \frac{\varepsilon}{2E\psi}\right) \left( 2F' \left( \psi \right) \right) d\psi \nonumber \\
%    &= \frac{2\varepsilon}{E} \int_{\psi_0}^{2\psi_0} \left( \psi - \frac{\gamma}{\varepsilon} \frac{1}{\psi}\right) d\psi \nonumber \\
%    &= \left( 3 - 2 \log 2 \right) \frac{\gamma}{E}.
%\end{align}
Defining a wave number as $k \coloneqq 2\pi/\lambda$, we have the energy dispersion relation in Eq.~(\ref{eq: strong nonlinearity}).
%\begin{equation}
%    E = \frac{3-2\log 2}{2\pi} \gamma k.
%\end{equation}
%Notably, despite the nonrelativistic nature of the original system, the strong nonlinearity gives rise to the emergent relativistic energy dispersion.

\begin{figure*}[t]
\centering
\includegraphics[width=1.0\linewidth]{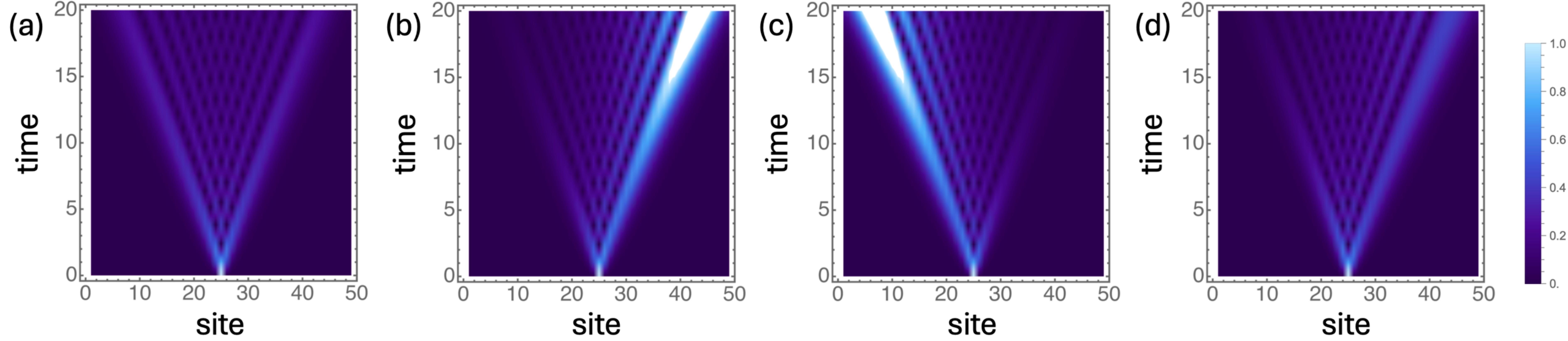} 
\caption{Dynamics of the nonlinear Hatano-Nelson model on the lattice under the open boundary conditions ($L=49$).
Plots of $\left| \psi_n \left( t \right) \right|$ as functions of site $n$ and time $t$ for (a)~$\gamma = 0$, $\varepsilon = 0$, (b)~$\gamma = 0.1$, $\varepsilon = 0$, (c)~$\gamma = -0.1$, $\varepsilon = 0$, and (d)~$\gamma = 0.1$, $\varepsilon = 0.5$.
The initial conditions are set as $\psi_n \left( t=0 \right) = \delta_{n, (L+1)/2}$.}	
    \label{afig: dynamics}
\end{figure*}

%%%%%%%%%%
\section{Nonlinear dynamics}

We investigate the dynamics of the nonlinear Hatano-Nelson model on the lattice by numerically solving the nonlinear Schr\"odinger equation $\ii \partial_t \ket{\psi} = \hat{H} \left( \ket{\psi} \right) \ket{\psi}$. 
In Fig.~\ref{afig: dynamics}, we present the time evolution of the wave function $\ket{\psi \left( t \right)} = \left( \psi_1 \left( t \right), \cdots, \psi_L \left( t \right) \right)^T$.
In the linear regime (i.e., $\varepsilon = 0$), the wave packet propagates unidirectionally toward the right (left) for $\gamma > 0$ ($\gamma < 0$), reflecting the non-Hermitian skin effect.
By contrast, in the nonlinear regime ($\varepsilon > 0$), the wave packet exhibits the bidirectional propagation.
This reciprocal dynamics is consistent with the nonlinearity-induced destabilization of the skin effect and the concomitant emergence of delocalized states via the Hopf bifurcation.

%%%%% Supplemental Material %%%%%
\clearpage
\widetext

\setcounter{secnumdepth}{3}

\renewcommand{\theequation}{S\arabic{equation}}
\renewcommand{\thefigure}{S\arabic{figure}}
\renewcommand{\thetable}{S\arabic{table}}
\setcounter{equation}{0}
\setcounter{figure}{0}
\setcounter{table}{0}
\setcounter{section}{0}
\setcounter{tocdepth}{0}

\numberwithin{equation}{section} 

\begin{center}
{\bf \large Supplemental Material for \\ \smallskip 
``Hopf Bifurcation of Nonlinear Non-Hermitian Skin Effect"}
\end{center}

%\tableofcontents

%%%%%%%%%%
\section{Existence and uniqueness of limit cycle}
    \label{asec: Lienard}

The existence and uniqueness of the limit cycle in the nonlinear Hatano-Nelson model are proved on the basis of the Li\'enard theorem~\cite{Jordan-textbook, Grimshaw-textbook, Perko-textbook}.
For $E, \psi \in \mathbb{R}$, the nonlinear Schr\"odinger equation reduces to the Li\'enard equation
\begin{equation}
    \frac{d^2\psi}{dx^2} + f \left( \psi \right) \frac{d\psi}{dx} + g \left( \psi \right) = 0
\end{equation}
with 
\begin{equation}
    f \left( \psi \right) \coloneqq 2 \left( \varepsilon \psi^2 - \gamma \right), \quad g \left( \psi \right) \coloneqq 2E\psi.
\end{equation}
Then, for $\gamma, \varepsilon, E > 0$, the conditions of the Li\'enard theorem are satisfied as follows:
\begin{itemize}
    \item For arbitrary parameters, $f$ and $g$ are continuously differentiable even and odd functions, respectively.

    \item For $\psi > 0$, we have $g \left( \psi \right) > 0$.

    \item Let us introduce the following odd function:
    \begin{equation}
        F \left( \psi \right) \coloneqq \int_0^\psi f \left( \chi \right) d\chi = 2 \left( \frac{\varepsilon}{3} \psi^3 - \gamma \psi \right).
    \end{equation}
    This function satisfies $\lim_{\psi\to \infty} F \left( \psi \right) = \infty$.
    Additionally, $F$ has exactly one positive root at $\psi = \sqrt{3\gamma/\varepsilon}$; 
    $F$ is negative for $0 < \psi < \sqrt{3\gamma/\varepsilon}$, and positive and monotonic for $\psi > \sqrt{3\gamma/\varepsilon}$.
\end{itemize}
Notably, the last condition is not satisfied for $\gamma \leq 0$.

%%%%%%%%%%
\section{Singular perturbation theory for weak nonlinearity}
    \label{asec: weak}

%%%%%%%%%%
\subsection{Continuum model}

We perturbatively derive a delocalized eigenstate associated with a limit cycle in phase space for weak nonlinearity $0 \leq \varepsilon \ll 1$.
We focus on the vicinity of the Hopf bifurcation and assume that the linear non-Hermitian parameter $\gamma$ is of the same order as the nonlinear non-Hermitian parameter $\varepsilon$, i.e., 
\begin{equation}
    \tilde{\gamma} \coloneqq \frac{\gamma}{\varepsilon} = \mathcal{O} \left( 1 \right).
\end{equation}
Under this assumption, the nonlinear Schr\"odinger equation reads
\begin{equation}
    \partial_x^2 \psi + 2 \varepsilon \left( \psi^2 - \tilde{\gamma} \right) \partial_x \psi + 2E \psi = 0.
        \label{aeq: Schrodinger weak nonlinear}
\end{equation}
To capture the $\mathcal{O} \left( \varepsilon^2 \right)$ correction of the energy dispersion, we here employ the Poincar\'e-Lindstedt method.
We reformulate Eq.~(\ref{aeq: Schrodinger weak nonlinear}) and obtain
\begin{equation}
    k^2 \partial_z^2 \psi + 2k\varepsilon \left( \psi^2 - \tilde{\gamma} \right) \partial_z \psi + 2E \psi = 0, \quad z \coloneqq kx.
        \label{aeq: Schrodinger Poincare-Lindstedt}
\end{equation}
Expanding both wave function $\psi$ and wave number $k$ as
\begin{equation}
    \psi = \psi_0 + \varepsilon \psi_1 + \varepsilon^2 \psi_2 + \mathcal{O} \left( \varepsilon^3 \right), \quad k = \sqrt{2E} + \varepsilon k_1 + \varepsilon^2 k_2 + \mathcal{O} \left( \varepsilon^3 \right),
\end{equation}
we have for $\mathcal{O} \left( 1 \right)$ and $\mathcal{O} \left( \varepsilon \right)$
\begin{align}
    &\partial_z^2 \psi_0 + \psi_0 = 0, 
        \label{aeq: PL1} \\
    &\partial_z^2 \psi_1 + \psi_1 = - \sqrt{\frac{2}{E}} \left( k_1 \left( \partial_z^2 \psi_0 \right) + \left( \psi_0^2 - \tilde{\gamma} \right) \partial_z \psi_0 \right).
        \label{aeq: PL2}
\end{align}

The solution to Eq.~(\ref{aeq: PL1}) with the initial condition $\psi_0 \left( z=0 \right) = 0$ is given by
\begin{equation}
    \psi_0 = r \sin z \quad \left( r \in \mathbb{R} \right).
\end{equation}
Substituting this into Eq.~(\ref{aeq: PL2}), we further have
\begin{align}
    \partial_z^2 \psi_1 + \psi_1 = - \sqrt{\frac{2}{E}} \left( - k_1 r \sin z + r \left( r^2 \sin^2 z - \tilde{\gamma} \right) \cos z \right).
        \label{aeq: PL2-1}
\end{align}
Since eliminating resonance terms imposes the conditions
\begin{equation}
    k_1 r = 0, \quad r \left( \frac{r^2}{4} - \tilde{\gamma} \right) = 0,
\end{equation}
the nontrivial solution follows as
\begin{equation}
    k_1 = 0, \quad r = 2\sqrt{\tilde{\gamma}}.
\end{equation}
Then, Eq.~(\ref{aeq: PL2-1}) is simplified to
\begin{equation}
    \partial_z^2 \psi_1 + \psi_1 = \frac{r^3}{2 \sqrt{2E}} \cos 3z = \frac{4\tilde{\gamma}^{3/2}
    }{\sqrt{2E}} \cos 3z,
\end{equation}
the solution of which with the boundary condition $\psi_1 \left( z=0 \right) = 0$ is
\begin{equation}
    \psi_1 = c \sin z + \frac{\tilde{\gamma}^{3/2}
    }{2\sqrt{2E}} \left( \cos z - \cos 3 z \right) \quad \left( c \in \mathbb{R} \right).
\end{equation}

Now, the nonlinear Schr\"odinger equation~(\ref{aeq: Schrodinger Poincare-Lindstedt}) for $\mathcal{O} \left( \varepsilon^2 \right)$ reads
\begin{equation}
    \partial_z^2 \psi_2 + \psi_2 = - \sqrt{\frac{2}{E}} \left( k_2 \partial_z^2 \psi_0 + 2\psi_0 \psi_1 \partial_z \psi_0 + \left( \psi_0^2 - \tilde{\gamma} \right) \partial_z \psi_1 \right).
\end{equation}
Eliminating resonance terms yields the conditions
\begin{equation}
    c=0, \quad k_2 = - \frac{\tilde{\gamma}^{2}
    }{4\sqrt{2E}}.
\end{equation}
Consequently, the approximate wave function is obtained as
\begin{align}
    \psi &\simeq \psi_0 + \varepsilon \psi_1 \nonumber \\
    &= 2\sqrt{\tilde{\gamma}} \sin z + \frac{\varepsilon \tilde{\gamma}^{3/2}
    }{2\sqrt{2E}} \left( \cos z - \cos 3z \right) \nonumber \\
    &\simeq 2 \sqrt{\tilde{\gamma}} \sin \left( 
 \left( \sqrt{2E} - \frac{\tilde{\gamma}^{2}
 }{4\sqrt{2E}} \varepsilon^2 \right) x\right) + \frac{\varepsilon \tilde{\gamma}^{3/2}
 }{2\sqrt{2E}} \left( \cos \sqrt{2E} x - \cos 3\sqrt{2E} x \right).
\end{align}
The corresponding energy dispersion follows as
\begin{equation}
    k = \sqrt{2E} - \frac{\tilde{\gamma}^{2}
    }{4\sqrt{2E}} \varepsilon^2 + \mathcal{O} \left( \varepsilon^3 \right), \quad \mathrm{i.e.}, \quad E = \frac{k^2}{2} + \frac{\tilde{\gamma}^{2}
    }{4} \varepsilon^2 + \mathcal{O} \left( \varepsilon^3 \right),
        \label{aeq: dispersion - continuum}
\end{equation}
which accompanies the $\mathcal{O} \left( \varepsilon^2 \right)$ correction due to nonlinear effects.

%%%%%%%%%%
\subsection{Lattice model}

In a similar manner to the continuum model, we investigate the lattice version of the nonlinear Hatano-Nelson model in the weakly nonlinear regime [i.e., $0 < \varepsilon \ll 1$, $\tilde{\gamma} \coloneqq \gamma/\varepsilon = \mathcal{O} \left( 1 \right)$] on the basis of the Poincar\'e-Lindstedt method.
Expanding the wave function and the modulation of the length scale as
\begin{equation}
    \ket{\psi} = \ket{\psi^0} + \varepsilon \ket{\psi^1} + \varepsilon^2 \ket{\psi^2} + \mathcal{O} \left( \varepsilon^3 \right), \quad 1 + \varepsilon a_1 + \varepsilon^2 a_2 + \mathcal{O} \left( \varepsilon^3 \right),
\end{equation}
we reformulate the nonlinear Schr\"odinger equation as
\begin{equation}
    \left( 1 + \varepsilon a_1 + \varepsilon^2 a_2\right)^2 \left( \psi_{n+1} + \psi_{n-1} - 2 \psi_{n} \right) + 2 \varepsilon \left( 1 + \varepsilon a_1 + \varepsilon^2 a_2\right) \left( \left| \psi_n \right|^2 - \tilde{\gamma} \right) \frac{\psi_{n+1} - \psi_{n-1}}{2} + 2E \psi_n = 0.
        \label{aeq: Schrodinger - lattice}
\end{equation}
At orders $\mathcal{O} \left( 1 \right)$ and $\mathcal{O} \left( \varepsilon \right)$, this yields
\begin{align}
    &\left( \psi_{n+1}^{0} + \psi_{n-1}^{0} - 2 \psi_{n}^{0} \right) + 2E \psi_n^{0} = 0, 
        \label{aeq: lattice0} \\
    &\left( \psi_{n+1}^{1} + \psi_{n-1}^{1} - 2 \psi_{n}^{1} \right) + 2E \psi_n^{1} = - 2a_1 \left( \psi_{n+1}^{0} + \psi_{n-1}^{0} - 2 \psi_{n}^{0} \right) - \left( \left| \psi_{n}^{0} \right|^2 - \tilde{\gamma} \right) \left( \psi_{n+1}^{0} - \psi_{n-1}^{0} \right).
        \label{aeq: lattice1}
\end{align}

The solution to Eq.~(\ref{aeq: lattice0}) with the boundary condition $\psi_0^0 = 0$ is given as
\begin{equation}
    \psi_n^0 = r_0 \sin k_0 n, \quad E \eqqcolon 1 - \cos k_0 \quad \left( r_0 \in \mathbb{R}, k_0 > 0 \right).
\end{equation}
Substituting this into Eq.~(\ref{aeq: lattice1}), we obtain
\begin{align}
    &\left( \psi_{n+1}^{1} + \psi_{n-1}^{1} - 2 \psi_{n}^{1} \right) + 2E \psi_n^{1} \nonumber \\
    &\quad = 4 a_1 r_0 \left( 1-\cos k_0 \right) \sin k_0 n - 2 r_0 \sin k_0 \left( r_0^2 \sin^2 k_0 n - \tilde{\gamma} \right) \cos k_0 n \nonumber \\
    &\quad = 4 a_1 r_0 \left( 1-\cos k_0 \right) \sin k_0 n - 2 r_0 \sin k_0 \left[ \left( \frac{r_0^2}{4} - \tilde{\gamma} \right) \cos k_0 n - \frac{r_0^2}{4} \cos 3k_0 n\right].
\end{align}
For $\tilde{\gamma} > 0$, eliminating resonance terms imposes the conditions
\begin{equation}
    a_1 = 0, \quad r_0 = 2 \sqrt{\tilde{\gamma}},
\end{equation}
yielding the solution
\begin{equation}
    \psi_n^1 = c \sin k_0 n + \frac{\tilde{\gamma}^{3/2}}{\sin 2k_0} \left( \cos k_0 n - \cos 3k_0 n \right) \quad \left( c \in \mathbb{R} \right)
\end{equation}
with the boundary condition $\psi_0^1 = 0$.

We proceed to the $\mathcal{O} \left( \varepsilon^2 \right)$ contribution of the original nonlinear Schr\"odinger equation,
\begin{align}
    &\left( \psi_{n+1}^{2} + \psi_{n-1}^{2} - 2 \psi_{n}^{2} \right) + 2E \psi_n^{2} \nonumber \\
    &\qquad = - 2a_2 \left( \psi_{n+1}^{0} + \psi_{n-1}^{0} - 2 \psi_{n}^{0} \right) - 2 \psi_n^0 \psi_n^1 \left( \psi_{n+1}^0 - \psi_{n-1}^0 \right) - \left( \left| \psi_n^0 \right|^2 - \tilde{\gamma} \right) \left( \psi_{n+1}^1 - \psi_{n-1}^1 \right).
\end{align}
Eliminating resonance terms on the right-hand side leads to
\begin{equation}
    a_2 = \frac{\tilde{\gamma}^2 \left( 3-4\cos^2 k_0 \right)}{8\cos k_0 \left( 1-\cos k_0\right)}, \quad c = 0.
\end{equation}

Consequently, the approximate wave function takes the form
\begin{align}
    \psi &\simeq \psi_0 + \varepsilon \psi_1 \nonumber \\
    &= 2\sqrt{\tilde{\gamma}} \sin \left[ k_0 \left( 1 + \frac{3-4\cos^2 k_0}{8\cos k_0 \left( 1-\cos k_0\right)} \left( \tilde{\gamma} \varepsilon \right)^2 \right) n\right] + \frac{\tilde{\gamma}^{3/2} \varepsilon}{\sin 2k_0} \left( \cos k_0 n - \cos 3k_0 n \right) \nonumber \\
    &=2\sqrt{\tilde{\gamma}} \sin \left[ k_0 \left( 1 + \frac{3-4\left( 1-E \right)^2}{8E \left( 1-E \right)} \left( \tilde{\gamma} \varepsilon \right)^2 \right) n\right] + \frac{\tilde{\gamma}^{3/2} \varepsilon}{\sin 2k_0} \left( \cos k_0 n - \cos 3k_0 n \right).
\end{align}
The corresponding energy dispersion relation is given by
\begin{equation}
    k = k_0 \left( 1 + \frac{3-4\left( 1-E \right)^2}{8E \left( 1-E \right)} \left( \tilde{\gamma} \varepsilon \right)^2 \right) + \mathcal{O} \left( \varepsilon^3 \right),
\end{equation}
further reducing to
\begin{equation}
    E = 1 - \cos k - \frac{k \sin k\left( 3-4\cos^2 k \right)}{8 \cos k \left( 1-\cos k \right)} \left( \tilde{\gamma} \varepsilon \right)^2 + \mathcal{O} \left( \varepsilon^3 \right).
\end{equation}
While this expression recovers Eq.~(\ref{aeq: dispersion - continuum}) for $\left| k \right| \ll 1$, it seems to break down for $k = \pm \pi/2$ as a consequence of lattice effects.

%%%%%%%%%%
\section{Nonlinear Hatano-Nelson model}
    \label{asec: HN}

%%%%%%%%%%
\subsection{Critical nonlinear skin effect in the continuum}

The nonlinear Schr\"odinger equation for the Hatano-Nelson model in the one-dimensional continuum reads
\begin{equation}
    \partial_x^2 \psi + 2 \left( \varepsilon \left| \psi \right|^2 - \gamma \right) \partial_x \psi + 2E \psi = 0.
        \label{asec: HN-NSE}
\end{equation}
We numerically solve this differential equation precisely at the Hopf bifurcation point $\gamma = 0$ [Fig.~\ref{afig: HN}\,(a)].
The wave function exhibits the algebraic decay characterized by $\left| \psi \right| \propto 1/\sqrt{x}$, consistent with the discussion presented in the main text.

%%%%%%%%%%
\subsection{Nonlinear skin effect on the lattice}

A discrete counterpart of Eq.~(\ref{asec: HN-NSE}) is given as
\begin{equation}
    \left( \psi_{n+1} + \psi_{n-1} - 2 \psi_{n} \right) + 2 \left( \varepsilon \left| \psi_n \right|^2 - \gamma \right) \frac{\psi_{n+1} - \psi_{n-1}}{2} + 2E \psi_n = 0,
\end{equation}
which can be rewritten as
\begin{equation}
  - \frac{1}{2} \left( 1-\gamma + \varepsilon \left| \psi_n \right|^2 \right) \psi_{n+1} - \frac{1}{2}  \left( 1 +\gamma - \varepsilon \left| \psi_n \right|^2 \right) \psi_{n-1} + \psi_n = E \psi_{n}.
\end{equation}
The corresponding nonlinear single-particle Hamiltonian is given by
\begin{equation}
    \hat{H} \left( \ket{\psi} \right) = \sum_{n} \left( - \frac{1 - \gamma + \varepsilon \left| \psi_n \right|^2}{2} \ket{n} \bra{n+1}
    - \frac{1 + \gamma - \varepsilon \left| \psi_{n+1} \right|^2}{2} \ket{n+1} \bra{n}
    + \ket{n} \bra{n} \right),
\end{equation}
where $\ket{\psi} = \left( \psi_1, \cdots, \psi_L \right)^T$ denotes a single-particle eigenstate, and $\ket{n}$ is a single-particle state localized at site $n$.

\begin{figure}[H]
\centering
\includegraphics[width=0.6\linewidth]{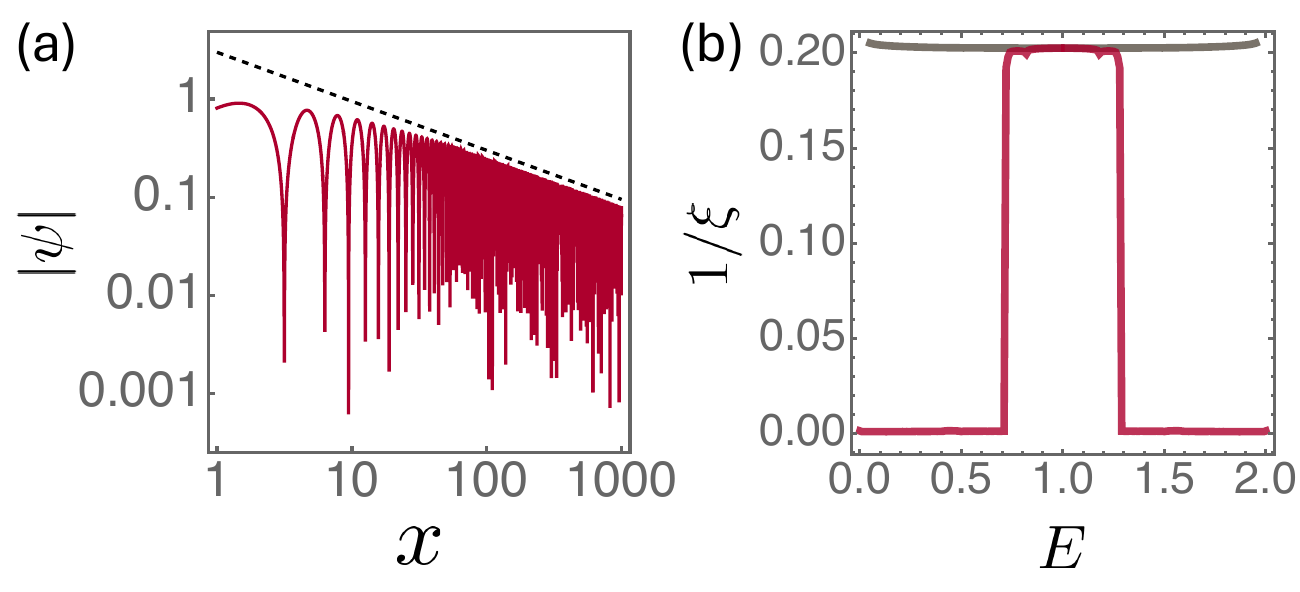} 
\caption{(a)~Wave function numerically obtained for the continuum nonlinear Hatano-Nelson model ($\gamma = 0$, $\varepsilon = 0.3$, $E = 0.5$) with the boundary conditions $\psi \left( x=0 \right) = 0$, $\psi' \left( x=0 \right) = 1$.
Dotted line: $\psi \left( x \right) = 3/\sqrt{x}$.
(b)~Inverse of the localization length, $1/\xi$, numerically obtained for the lattice nonlinear Hatano-Nelson model ($L=1000$, $\gamma = 0.2$, $\varepsilon = 0.3$) as a function of single-particle energy $E$.}	
    \label{afig: HN}
\end{figure}

In Fig.~\ref{afig: HN}\,(b), we show the dependence of the localization length $\xi$ on the single-particle energy $E$.
In the linear regime (i.e., $\varepsilon = 0$), the localization length $\xi = 1/\gamma$ of the skin state remains constant and exhibits no dependence on $E$.
On the other hand, in the nonlinear regime, the system develops mobility edges at $E \simeq 0.7, 1.3$;
while $\xi$ is finite within the energy window $0.7 \lesssim E \lesssim 1.3$, it diverges for $E \lesssim 0.7$, $E \gtrsim 1.3$.
The presence of these mobility edges is reminiscent of those in the original (i.e., disordered) Hatano-Nelson model~\cite{Hatano-Nelson-96, *Hatano-Nelson-97}.

%%%%%%%%%%
\section{Nonlinear non-Hermitian Su-Schrieffer-Heeger model}
    \label{asec: SSH}

As another prototypical example exhibiting the nonlinear skin effect, we consider a nonlinear non-Hermitian extension of the Su-Schrieffer-Heeger model~\cite{SSH-79}:
\begin{align}
    \hat{H} \left( \ket{\psi} \right) &= \sum_{n} \left[ \left( v+\gamma - \varepsilon\,( \left| a_n \right|^2 + \left| b_n \right|^2 ) \right) \ket{n, {\rm B}} \bra{n, {\rm A}} + \left( v-\gamma + \varepsilon\,( \left| a_n \right|^2 + \left| b_n \right|^2 ) \right) \ket{n, {\rm A}} \bra{n, {\rm B}} \right. \nonumber \\
    &\qquad\qquad\qquad\qquad\qquad\qquad\qquad\qquad\qquad  + w \left( \ket{n+1, {\rm A}} \bra{n, {\rm B}} + \ket{n, {\rm B}} \bra{n+1, {\rm A}} \right) \Big],
\end{align}
where $\ket{n, {\rm A}}$ and $\ket{n, {\rm B}}$ denote single-particle states localized at site $n$ with sublattices ${\rm A}$ and ${\rm B}$, respectively, and $\ket{\psi} = \left( a_1, b_1, \cdots, a_L, b_L \right)^T$ represents a single-particle eigenstate.
Here, $v \in \mathbb{R}$ and $w \in \mathbb{R}$ are the intracell and intercell hopping amplitudes, respectively;
$\gamma \in \mathbb{R}$ characterizes the nonreciprocal hopping amplitude, and $\varepsilon \geq 0$ is its nonlinear modulation, analogous to the nonlinear Hatano-Nelson model.
In the absence of nonlinearity (i.e., $\varepsilon = 0$), $\hat{H}$ reduces to a canonical non-Hermitian model that hosts the skin effect~\cite{Lee-16, YW-18-SSH}, whose Bloch Hamiltonian is given by
\begin{equation}
    H \left( k \right) = \left( v+w \cos k \right) \sigma_x + \left( w\sin k + \ii \gamma \right) \sigma_y 
    = \begin{pmatrix}
        0 & v+\gamma + w e^{-\ii k} \\
        v-\gamma + w e^{\ii k} & 0
    \end{pmatrix}.
\end{equation}

The single-particle Schr\"odinger equation $\hat{H} \left( \ket{\psi} \right) \ket{\psi} = E \ket{\psi}$ reads
\begin{align}
    \left( v-\gamma + \varepsilon\,( \left| a_n \right|^2 + \left| b_n \right|^2 ) \right) b_n + wb_{n-1} &= E a_n, \\
    \left( v+\gamma - \varepsilon\,( \left| a_n \right|^2 + \left| b_n \right|^2 ) \right) a_n + wa_{n+1} &= E b_n,
\end{align}
which can be recast as
\begin{equation}
    \begin{pmatrix}
        a_{n+1} \\ b_{n+1}
    \end{pmatrix} = T \left( a_n, b_n \right) \begin{pmatrix}
        a_{n} \\ b_{n}
    \end{pmatrix}
\end{equation}
with the nonlinear transfer matrix $T \left( a_n, b_n \right)$ defined by
\begin{equation}
    T \left( a_n, b_n \right) \coloneqq \begin{pmatrix}
       -\cfrac{v+ \gamma_{\varepsilon}}{w} & \cfrac{E}{w} \\
       - \cfrac{E \left( v+\gamma_{\varepsilon} \right)}{w \left( v-\gamma_{\varepsilon} \right)} & \cfrac{E^2 - w^2}{w \left( v-\gamma_{\varepsilon} \right)}
    \end{pmatrix}, \quad \gamma_{\varepsilon} \left( a_n, b_n \right) \coloneqq \gamma - \varepsilon \left( \left| a_n \right|^2 + \left| b_n \right|^2 \right).
\end{equation}
The spatial distribution of the wave function for given energy $E$ is then determined via the successive application of the nonlinear transfer matrix $T \left( a_n, b_n \right)$.

\begin{figure}[H]
\centering
\includegraphics[width=1.0\linewidth]{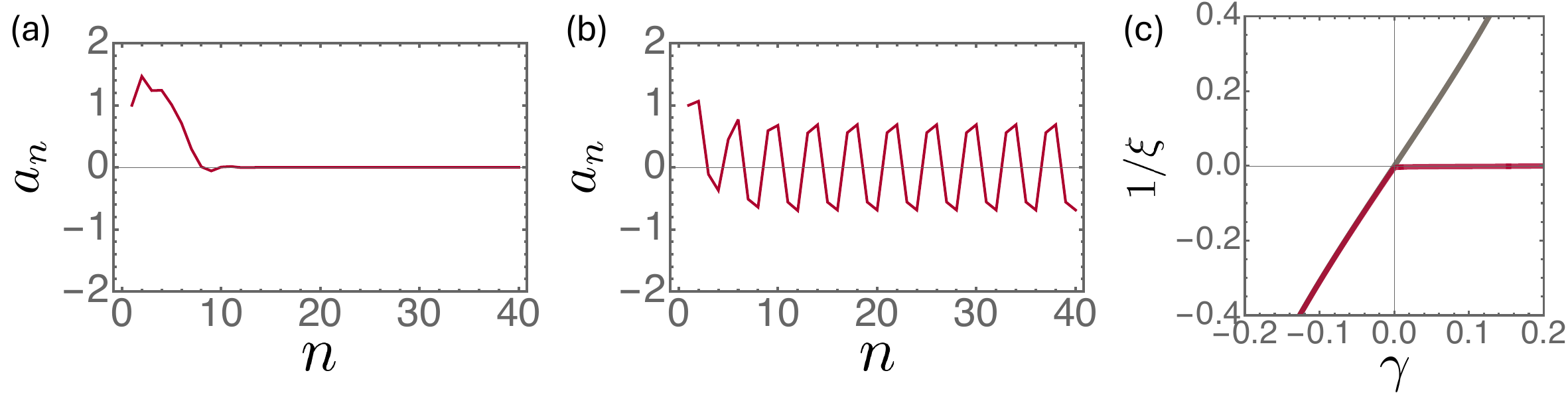} 
\caption{Nonlinear non-Hermitian Su-Schrieffer-Heeger model ($v=1/3$, $w=1$, $E=1$).
Boundary conditions: $a_1 = b_1 = 1$.
(a, b)~Wave function for $\varepsilon = 0.3$ with (a)~$\gamma = -0.2$ and (b)~$\gamma = 0.2$.
(c)~Inverse of the localization length, $1/\xi$ ($\left| a_n \right| \sim e^{n/\xi}$), numerically obtained over $500 \leq n \leq 1000$ [$\varepsilon = 0$ (gray), $\varepsilon = 0.3$ (red)].}	
    \label{afig: SSH}
\end{figure}

In the presence of nonlinearity $\varepsilon > 0$, the wave function exhibits qualitatively distinct behavior depending on the sign of $\gamma$:
it decays toward the right for $\gamma < 0$ [Fig.~\ref{afig: SSH}\,(a)], whereas it oscillates for $\gamma > 0$ [Fig.~\ref{afig: SSH}\,(b)].
This behavior is similar to that of the nonlinear Hatano-Nelson model and signals the Hopf bifurcation of the nonlinear skin effect.
In Fig.~\ref{afig: SSH}\,(c), we further show the localization length as a function of linear non-Hermiticity $\gamma$. 
In the linear regime, the localization length remains finite for arbitrary $\gamma \neq 0$, manifesting the skin effect.
In the nonlinear regime, by contrast, the localization length diverges for $\gamma > 0$, implying the delocalization due to the emergence of a limit cycle in the associated phase space.

%\bibliography{NH_top.bib}

\end{document}